\documentclass[prb,aps,twocolumn,amsmath,amssymb,floatfix,
superscriptaddress]{revtex4}
\usepackage[dvips]{graphics}
\usepackage{color}
\definecolor{dred}{rgb}{0,0,0.6}
\usepackage{graphicx}  
\usepackage{dcolumn}   
\usepackage{bm}        
\usepackage{amssymb}   
\usepackage{amsmath}
\usepackage{braket}
\usepackage[mathscr]{euscript}
\usepackage{color}
\usepackage{hyperref}
\begin{document}

\author{Kallol Mondal}
\email[E-mail: ]{kallolsankarmondal@gmail.com}
\affiliation{Physics and Applied Mathematics Unit, Indian Statistical Institute, 203 Barrackpore Trunk Road, Kolkata-700108, India}

\author{Sudin Ganguly}
\email[E-mail: ]{sudinganguly@gmail.com}
\affiliation{Physics and Applied Mathematics Unit, Indian Statistical Institute, 203 Barrackpore Trunk Road, Kolkata-700108, India}

\author{Santanu K. Maiti}
\email[E-mail: ]{santanu.maiti@isical.ac.in}
\affiliation{Physics and Applied Mathematics Unit, Indian Statistical Institute, 203 Barrackpore Trunk Road, Kolkata-700108, India}

\date{\today}
\title{Spin-dependent transport in a driven noncolinear antiferromagnetic fractal network}
\begin{abstract}
Noncolinear magnetic texture breaks the spin-sublattice symmetry which gives rise to a spin-splitting effect. Inspired by this, we study the spin-dependent transport properties in a noncolinear antiferromagnetic fractal structure, namely, the Sierpinski Gasket (SPG) triangle. We find that though the spin-up and spin-down currents are different, the degree of spin polarization is too weak. Finally, we come up with a proposal, where the degree of spin polarization can be enhanced significantly in the presence of a time-periodic driving field. Such a prescription of getting spin-filtering effect from an unpolarized source in a fractal network is completely new to the best of our knowledge. Starting from a higher generation of SPG to smaller ones, the precise dependencies of driving field parameters, spin-dependent scattering strength, interface sensitivity on spin polarization are critically investigated. The spatial distribution of spin-resolved bond current density is also explored. Interestingly, our proposed setup exhibits finite spin polarization for different spin-quantization axes. Arbitrarily polarized light is considered and its effect is incorporated through Floquet-Bloch ansatz. All the spin-resolved transport quantities are computed using Green's function formalism following the Landauer-B\"{u}ttiker prescription. The present work brings forth new insights into spintronic properties of noncolinear antiferromagnetic SPG and should entice the AFM spintronic community to explore other fractal structures with the possibility of unconventional features. 
\end{abstract}
\maketitle
\section{\label{sec:intro}Introduction}
Antiferromagnetic (AFM) spintronics~\cite{afm-rev1,afm-rev2,afm-rev3}, a new paradigm has been emerged in the field of spintronics owing to its intriguing superiority over the conventional ferromagnetic spin-based devices. Due to the net vanishing magnetic moment, AFM materials are robust against magnetic field perturbation, do not produce any stray field, and display ultrafast magnetization dynamics~\cite{fast1,fast2}. These features enable the antiferromagnets as promising candidates in future spintronic applications.

Depending upon the orientation of the magnetic moments, AFM materials can be of two kinds -- colinear and noncolinear. The colinear AFM materials, due to the symmetry of the spin sublattices, in general, do not produce spin-polarized currents~\cite{afm-rev3}. However, spin sublattice symmetry can be broken by means of external perturbations and AFM materials show impressive performance in spin transport. For instance, tailored layered structures comprising of ferromagnets, antiferromagnets, and ferro/non-magnets exhibit to transport spin currents~\cite{layer1,layer2,layer3,layer4}, circularly polarized subterahertz irradiation enables spin pumping effect in insulating colinear antiferromagnet~\cite{subtera}, generation of spin-polarized current by applying a bias voltage in colinear antiferromagnetic insulator~\cite{biasv}, etc. On the other hand, as the spin sublattice symmetry is no longer preserved, various fascinating features, such as anomalous Hall effect~\cite{ahe1,ahe2,ahe3,ahe4,ahe5}, spin Hall effect~\cite{she1}, inverse spin Hall effect~\cite{she1,she2,she3}, anomalous Nernst effect~\cite{ane}, etc., have been observed in noncolinear AFM materials. Recently, it has been shown that an electrical current can be polarized in noncolinear antiferromagnets as a consequence of the symmetry breaking~\cite{jakub1} and that is analogous to the spin-polarized current in ferromagnets. Given the possibility that a noncollinear magnetic structure can produce spin-polarized current analogous to the spin-orbit coupled systems~\cite{jakub1,jakub2}, we wish to explore the spin-dependent transport in a fractal structure, namely the Sierpinski Gasket (SPG) triangle. 

SPG structure, like the other fractals, falls under the category somewhat in between perfectly ordered and completely disordered systems, where finite ramification and self-similarity induce the localization. SPG structures exhibit Cantor set energy spectrum and highly degenerate localized states~\cite{domany}, which becomes delocalized in the presence of a magnetic field~\cite{banavar}. Numerous efforts have been made so far to study the electronic properties of the SPG structures under different scenarios and several other unique features have been observed~\cite{rammal,gordon,wasch,wang1,wang2,mayer,maiti-prb-spg,veen1,veen2,ili}. However, the study of spin transport in SPG is rare and therefore, certainly desirable for understanding the behavior in self-similar geometries having multiple loops and also for future spintronic applications. 

Thanks to the recent advancement of fabrication techniques, SPG structures have been realized experimentally with different materials, such as submicrometer-width Al wires~\cite{gordon-prl}, aromatic compounds~\cite{js-natchem}, metal-organic compounds~\cite{li-acs}, and by manipulation on CO molecules of Cu(111) surface~\cite{snk}, by deposition of Bi on InSb(111)B~\cite{chen-liu}. At the same time, the possibility of tailoring magnetic textures~\cite{mag-tex1,mag-tex2} engenders the present work more compelling in the field of AFM spintronics.  

By assuming the noncolinear arrangement of the magnetic moments in the SPG with the net-zero moment, we find that the behaviors of the opposite spin transmission probabilities are different from each other. For a magnetic materials, pin-dependent scattering takes place when itinerant electrons interact with local magnetic moments. But, the fact is that for a magnetic system with vanishing net magnetization it is usually very hard to get spin polarization. We establish that a noncolinear arrangement can provide a finite spin-filtration and the efficiency can be further improved at large degree (more than $90\%$) once we irradiate the sample. The spectral peculiarity, fractal-like gapped energy spectrum, and the coexistence of both conducting and localized states play the central role of getting non-trivial signatures in antiferromagnetic fractal lattices in presence of light. These features are usually absent in completely perfect or fully uncorrelated (random) disordered lattices. This prescription is completely new, to the best of our knowledge. Additionally, we detect multiple spin-dependent mobility edges which makes the SPG a promising candidate in future spintronic applications.

The effect of irradiation is incorporated through the standard Floquet-Bloch ansatz in the minimal coupling scheme~\cite{kallol-scirep,gomez-prl,sambe,grifoni,lght1,lght2, sudin-carbon, sudin-jap, maiti-prb2020}. The spin-dependent two-terminal transmission probabilities are computed using the well-known Green's function formalism, based on Landauer-B\"{u}ttiker prescription~\cite{etms,qtat}. The mobility edges are identified by superposing the total density of states (DOS) and the spin-dependent transmission probabilities.

The salient features obtained in the present work are: (i) appearance of a certain fraction of conducting states due to noncolinear magnetic texture, (ii) small but finite spin polarization associated with multiple spin-dependent mobility edges in the AFM fractal lattice, (iii) significant enhancement of the degree of polarization (more than $90\%$) in the presence of a driving field. Our analysis may help in designing future spin-based devices at the nanoscale level and to study some fascinating phenomena in similar kinds of AFM fractal lattices and other AFM topological systems.

The rest of the work is organized as follows. In Sec. II, we present our model quantum system and the Hamiltonian in the presence of an arbitrarily polarized light having a noncolinear arrangement of the magnetic moments with a zero net moment. In this section, we also present a detailed theoretical description for the calculations of spin-resolved two-terminal transmission probability, currents, spin polarization coefficient, and spin-resolved bond current density. All the results are critically investigated in Sec. III. Finally, in Sec. IV, we conclude our essential findings.


\section{\label{sec:formalism}Quantum system and theoretical formulation}
\subsection{SPG triangle and the Hamiltonian}
\begin{figure}[ht!]
\centering
\includegraphics[width=0.45\textwidth]{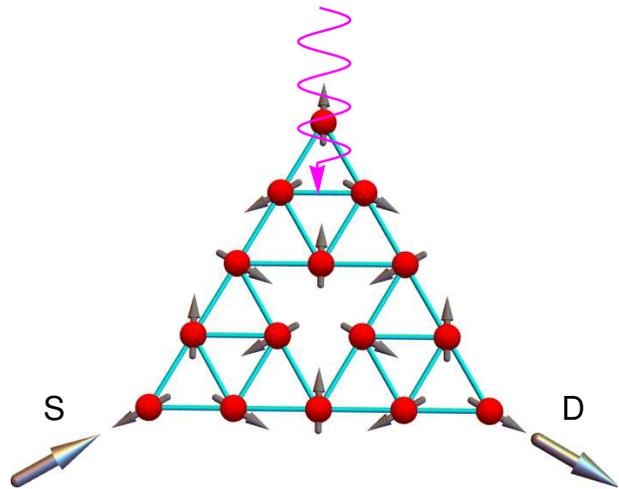} 
\caption{(Color online). Schematic view of an irradiated 3rd generation SPG fractal network with noncolinear AFM texture. The atomic sites are located at the vertices of each equilateral triangle, as shown by red solid spheres. The direction of the moment at each site is denoted with dark grey arrow. The SPG is attached to two electrodes (source $S$ and drain $D$).}
\label{spg}
\end{figure}

Figure~\ref{spg} depicts the schematic of our proposed device where an SPG network is connected with two one dimensional (1D) semi-infinite leads, namely source ($S$) 
and drain ($D$). An unpolarized electron beam is injected from the source, and the electrons are allowed to pass through the SPG network. We assume that the local moments of the network are arranged antiferromagnetically and the SPG is irradiated with an arbitrarily polarized light. Under such a scenario, the tight-binding Hamiltonian describing the SPG network can be expressed in the following form~\cite{afm-ham1,afm-ham2,afm-ham3}
\begin{equation}
H_{\text{SPG}} = \sum_n \epsilon_n \mathbf{c}_n^\dagger \mathbf{c}_n - \mathcal{J}\sum_n \mathbf{c}_n^\dagger \mathbf{S}_n \cdot \boldsymbol{\sigma} \mathbf{c}_n + \sum_{\langle nm \rangle}  \tilde{t}_{nm}\mathbf{c}_n^\dagger  \mathbf{c}_m.
\label{ham}
\end{equation}

The first term is associated with the on-site energy, where $\mathbf{c}_n^\dagger = (c_{n\uparrow}^\dagger, c_{n\downarrow}^\dagger)$, and $\uparrow, \downarrow$ refer to the spin projection along the quantization axis. $\epsilon_n$ is the on-site energy at the $n$-th site. 

The second term denotes the exchange interaction between the local magnetic moments and the conduction electron spin. $\mathbf{S}_n$ is the local moment at site $n$ and ${\mathcal J}$ is the exchange interaction strength. In the presence of the magnetic moments, a spin dependent scattering (SDS) factor~\cite{sds1,sds2} ${\mathbf h}_n = {\mathcal J}\mathbf{S}_n$ appears due to the interaction of incoming electrons with the magnetic moments. The strength of the SDS parameter $|{\mathbf h}|$ (we set ${\mathbf h_n}={\mathbf h}$, for all $n$), is assumed to be isotropic. $\boldsymbol{\sigma}$ is the Pauli spin vector and $\boldsymbol{\sigma} = (\sigma_x, \sigma_y, \sigma_z)$. {\it Here the alignment of the moments are assumed in such a way that the net magnetic moment at each triangular plaquette becomes zero}. Let us consider the topmost triangle in Fig.~\ref{spg}. If we denote the moment at the top vertex by $\mathbf{S}_1$, then following the clock-wise convention, the orientations of the three magnetic moments in that triangle in Cartesian coordinate can be written as
\begin{subequations}
 \begin{eqnarray}
 \mathbf{S}_1 & = & \langle S\rangle ~\hat{\bf y} \\
 \mathbf{S}_2 & = & \hat{R}(\hat{\bf z}, - \frac{2 \pi}{3})\cdot \mathbf{S}_1 \\
 \mathbf{S}_3 & = & \hat{R}(\hat{\bf z}, - \frac{4 \pi}{3})\cdot \mathbf{S}_1 
 \end{eqnarray}
  \label{eq:spin-mag}
\end{subequations}
where, $\hat{R}(\hat{\bf z},\theta)$ is the rotation operator that rotates a vector by an angle $\theta$ about the $z$-axis (perpendicular to the lattice plane) and $\langle S\rangle$ is the magnitude of the spin vector.  Clearly, the net magnetic moment  $\mathbf{S}_1 +\mathbf{S}_2 +\mathbf{ S}_3=0$. Following the same prescription, we arrange all other magnetic moments at different lattice sites, and the resultant magnetic moment of the SPG triangle becomes zero. We shall investigate the spin-filtration efficiency through such a noncolinear antiferromagnetic fractal structure. 

The last term in Eq.~\ref{ham} is associated with nearest-neighbor hopping (NNH) of electrons. $\tilde{t}_{nm}$ is the NNH integral in the presence of light, where $(n,m)$ are the site indices. When the system is irradiated with light, the Hamiltonian becomes time-dependent and complicated as well. With the help of Floquet-Bloch ansatz~\cite{sambe,grifoni,gomez-prl}, the effect of light irradiation is incorporated through the vector potential $\mathbf{A}(\tau)$. With the Peierls substitution, the vector potential $\mathbf{A}(\tau)$ is introduced through a phase $\frac{e}{c\hbar}\int {\mathbf A}(\tau)\cdot d{\mathbf l}$, where the symbols $e$, $c$, and $\hbar$ carry their usual meaning. Without the loss of any generality, the vector potential can be taken in the form $$\mathbf{A(\tau)} = (A_x \sin(\Omega \tau), A_y \sin(\Omega \tau + \phi),0),$$ which represents an arbitrarily polarized field in the $x$-$y$ plane. $A_x$ and $A_y$ are the field amplitudes, and $\phi$ is the phase. Depending upon the choices of $A_x, A_y$ and $\phi$, we get different polarized (circular, elliptic and linear) lights. In the presence of such irradiation, the effective hopping integral renormalizes as~\cite{gomez-prl,lght1,lght2}
\begin{equation}
\tilde{t}_{nm} \rightarrow t_{nm}^{pq} = t_{nm} \times \frac{1}{\mathbb{T}}\int_0^\mathbb{T} e^{i \Omega \tau(p-q)} e^{i \mathbf{A}(\tau) \cdot \mathbf{d}_{nm}} d\tau
\label{effhop}
\end{equation}
where $\mathbf{d}_{nm}$ is the vector joining the nearest-neighbor sites, $t_{nm}$ is the NNH strength in the absence of light, and, is assumed to be isotropic that is $t_{nm}=t$. The superscripts $p$ and $q$ are integers and correspond to the Floquet band indices.  $\mathbb{T}$ and $\Omega$ be the time period and frequency of the driving field. Here the vector potential is expressed in units of $ea/c\hbar$ ($a$ being the lattice constant, is taken to be 1\AA, for simplification).  

Assuming, $\mathbf{d}_{nm}=d_x\hat{\bf x} + d_y\hat{\bf y}$ and using the explicit form of $\mathbf{A(\tau)}$, Eq.~\ref{effhop} can further be written as 
\begin{eqnarray}
t_{nm}^{pq} &=&  \frac{t}{\mathbb{T}}\int_0^\mathbb{T} e^{i \Omega \tau(p-q)} e^{iA_xd_x\sin{\Omega\tau}} e^{iA_yd_y\sin{\left(\Omega\tau +\phi\right)}} d\tau\nonumber\\
&=&t e^{i (p-q)\Theta } J_{(p-q)}\left(\Gamma\right) 
\label{effhop1}
\end{eqnarray}
where, 
\begin{eqnarray}
\Gamma &=& \sqrt{\left(A_xd_x\right)^2 + \left(A_yd_y\right)^2 + 2A_xA_yd_xd_y\cos{\phi}}\\
\Theta &=& \tan^{-1}{\left(\frac{A_yd_y\sin{\phi}}{A_xd_x + A_yd_y\cos{\phi}}\right)}.
\end{eqnarray}
$J_{(p-q)}$ is the $(p-q)$-th order Bessel function of the first kind. From Eq.~\ref{effhop1}, it is now evident that the effective NNH integral depends on the hopping direction. Therefore, a spatial anisotropy can be achieved in the irradiated SPG network. 

The Floquet bands are the characteristic manifestation of the time-periodic driving field. A periodically driven ${\mathbb D}$-dimensional lattice is mapped to an undriven ${\mathbb D}+1$-dimensional lattice~\cite{gomez-prl,lght1}. Under such condition, the initial Bloch band splits into Floquet-Bloch (FB) bands, where the coupling between FB bands depends directly on the driving frequency regime. This undriven ${\mathbb D}+1$ dimensional lattice can be thought of as if the SPG is connected to its several virtual copies arranged vertically to the lattice plane. In the high-frequency limit, the Floquet bands decoupled from each other, and only the zeroth-order Floquet band ($p=q=0$) dominates over the other higher-order terms in $p$ and $q$ in Eq.~\ref{effhop1}. Because of this, the coupling between the parent SPG lattice and its virtual copies becomes vanishingly small. This scenario is no longer valid in the low-frequency regime, where the virtual copies are directly coupled to the parent SPG lattice. Therefore, in the low-frequency limit, several virtual copies of the SPG lattice come into the picture. Consequently, the effective size of the system increases. This could decrease the spin-relaxation length. 

In view of the above analysis, we restrict the present analysis to the high-frequency limit without loss of any generality. The high-frequency limit is valid as long as the frequency is larger than the bandwidth of the undriven system~\cite{lght1}, that is $\hbar\omega > 4t$. Assuming the NNH integral $t=1\,$eV, in this limiting case, the light frequency should be about $10^{15}\,$Hz, which lies in the near-ultraviolet regime. The electric field associated with this high-frequency light is $\sim 10^4\,$V/m, and the magnetic field is $\sim 10^{-5}\,$T. Since the magnetic field associated with the light irradiation is vanishingly small, any Zeeman-like interaction will not have any noticeable spin-splitting effect and thus its effect can safely be ignored. The intensity of the light irradiation is $\sim 10^5\,$W/m$^2$. Such light intensities or even higher intensities have been used in several other recent works~\cite{high-frq,cwd1,cwd2}.

\subsection{Formulation of spin-dependent transmission probabilities, transport currents, spin polarization, and bond current densities}
\subsubsection{Spin-dependent transmission probabilities}
To study the spin-dependent transport phenomena, we have employed Green's function formalism~\cite{etms,qtat,land,tr1,tr2}.  The characteristic features of spin-dependent transport can be understood from the behavior of the two-terminal spin-dependent transmission functions. The spin-dependent transmission probability $T_{\sigma \sigma^\prime}$ of an incoming electron with spin $\sigma$ being transmitted through the SPG network and collected at the drain with spin $\sigma^\prime$ is given by~\cite{etms,qtat,land}
\begin{equation}
T_{\sigma \sigma^\prime} =  \text{Tr}\left[ \Gamma_S^\sigma \mathcal{G}^r \Gamma_D^{\sigma^\prime} \mathcal{G}^a \right],
\end{equation}
where $\Gamma_{S(D)}^{\sigma(\sigma^\prime)} = - 2 \text{Im}[\Sigma_{S(D)}^{\sigma(\sigma^\prime)}]$ are the coupling matrices~\cite{tr1}. $\Gamma_{S(D)}^{\sigma}$ is the contact self-energy due to the source (drain) lead. The matrices $\mathcal{G}^a$ and $\mathcal{G}^r$ are the advanced and retarded Green's functions, respectively. $\mathcal{G}^r = (\mathcal{G}^a)^\dagger$. We must mention that, if $\sigma = \sigma^\prime$, then we get pure transmission and for $\sigma \neq \sigma^\prime$, we get the spin-flip transmission. 

Now, using the tight-binding Hamiltonian and the self-energy matrices, we construct the retarded Green's function matrix as following~\cite{etms,qtat,land,tr1,tr2}
\begin{equation}
\mathcal{G}^r = \left( E - H_{\text{SPG}}- \Sigma_S - \Sigma_D \right)^{-1}
\end{equation}
where $E$ is the energy of the incoming electrons.  

The net up and down spin transmission probabilities are defined in the following way
\begin{equation}
T_\uparrow =  T_{\uparrow \uparrow} +   T_{\downarrow \uparrow} ~~~~ \text{and}~~~~  T_\downarrow =  T_{\downarrow\downarrow} +   T_{\uparrow\downarrow} .
\end{equation}

\subsubsection{Spin-dependent transport currents and spin polarization}
 Once we compute the spin-dependent transmission probabilities, the calculation of different spin-dependent currents is quite viable. We compute the spin-dependent current at absolute zero temperature by integrating the spin-dependent transmission probability over a specified energy window associated with the bias voltage. The spin-dependent current $I_\sigma$ is expressed as~\cite{etms,qtat,land}, 
\begin{equation}
I_{\sigma } = \frac{e}{h}\int_{E_F - \frac{eV}{2}}^{E_F + \frac{eV}{2}} T_{\sigma }(E)\, dE 
\label{def-curr}
\end{equation}
where $\sigma=\uparrow,\downarrow$, $e$ and $h$ are the electronic charge and Planck's constant respectively. $E_F$ denotes the equilibrium Fermi energy. Assuming the broadening of energy levels due to the coupling of the SPG network and the contact leads is larger than the thermal broadening $k_BT$, we can safely ignore the thermal contribution in this analysis. 

So far, the spin-dependent transmission probabilities and currents are discussed assuming that the spin-quantization axis is along the $z$-direction. Whenever, in the present work, we discuss the spin-resolved transmission probabilities and currents, the spin-quantization axis is assumed to be along the $z$-direction. 
However, since the noncolinear arrangement of the moments breaks the spin-sublattice symmetry, it is expected that all the three components, namely, $x$, $y$, and $z$-components of the spin-polarized currents should be finite. Therefore, we compute the spin-polarized current for all three components.

The $\alpha$-component of the spin-polarized current can be evaluated with the following expression
\begin{equation}
I_\alpha^s = \frac{e}{h}\int_{E_F - \frac{eV}{2}}^{E_F + \frac{eV}{2}} T_\alpha^s(E)\, dE
\label{spin-curr}
\end{equation}
where $T_\alpha^s$ is the $\alpha$-component of the spin polarized transmission coefficient and is calculated as~\cite{po-chang,riv}
\begin{equation}
T_\alpha^s =  \text{Tr}\left[\sigma_\alpha \Gamma_S \mathcal{G}^r \Gamma_D \mathcal{G}^a \right].
\label{tspin}
\end{equation}
Here $\sigma_\alpha=\sigma_x,\sigma_y,\sigma_z$, are the Pauli matrices.

Finally, we define the spin polarization coefficient as the ratio between the spin-polarized current and the total charge current as~\cite{rai-prb-2012,mp}
\begin{equation}
P_\alpha = \frac{I_\alpha^s}{I}
\end{equation}
where $I_\alpha^s$ $(\alpha=x,y,z)$ denotes the $\alpha$-component of the spin-polarized current and $I$ is the total charge current. The total charge current can be computed by replacing $\sigma_\alpha$ with an identity matrix in Eq.~\ref{tspin}. $P_\alpha$ can take values between $-1$ to $ 1$. $P_\alpha = 0$ implies zero polarization, whereas $P_\alpha = 1\, (-1)$ indicates $100 \%$ spin polarization, corresponding to up (down) spin current.

\begin{figure*}[t!]
\centering
\includegraphics[width=0.45\textwidth]{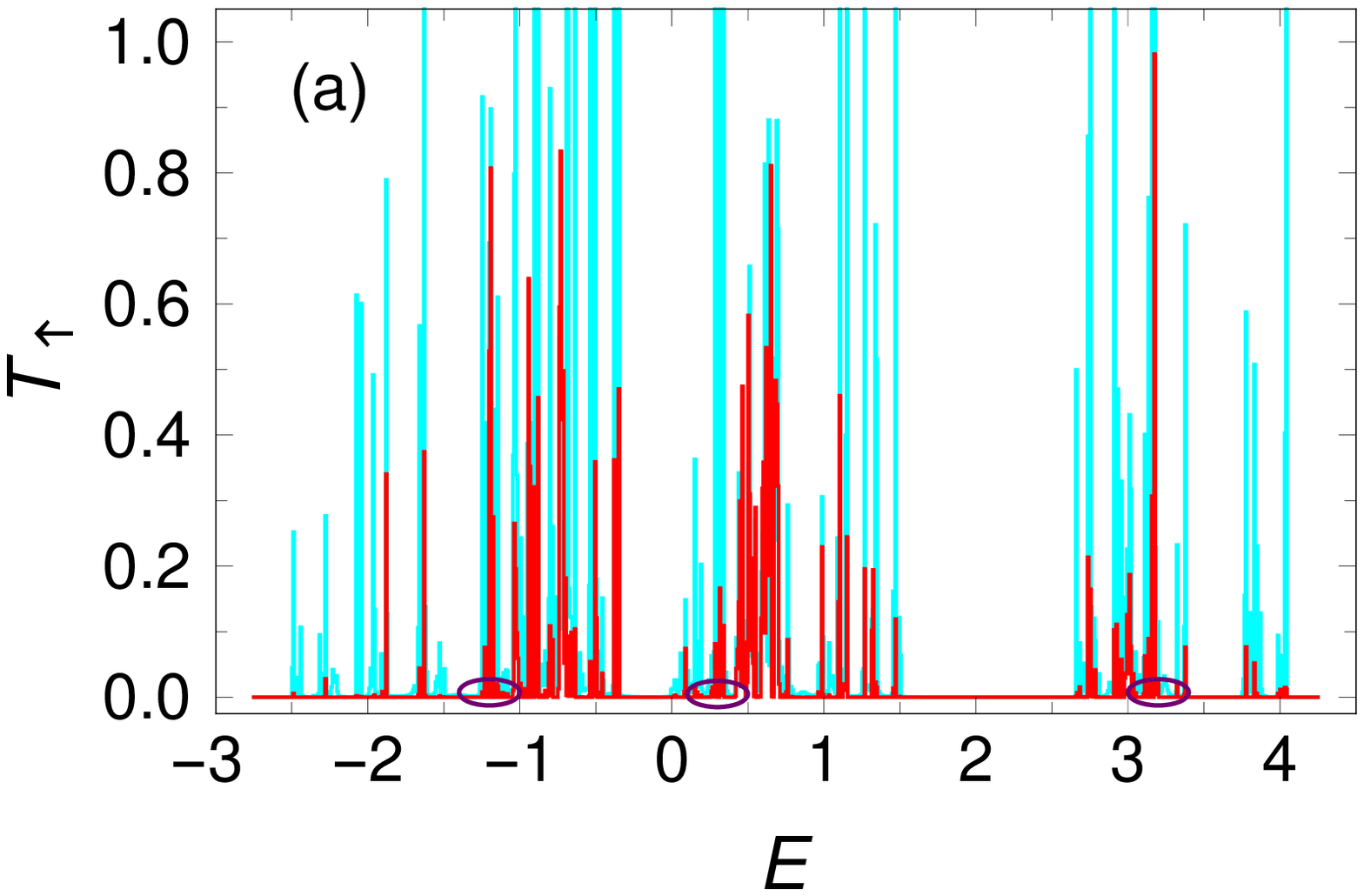} \hspace{0.25cm}
\includegraphics[width=0.45\textwidth]{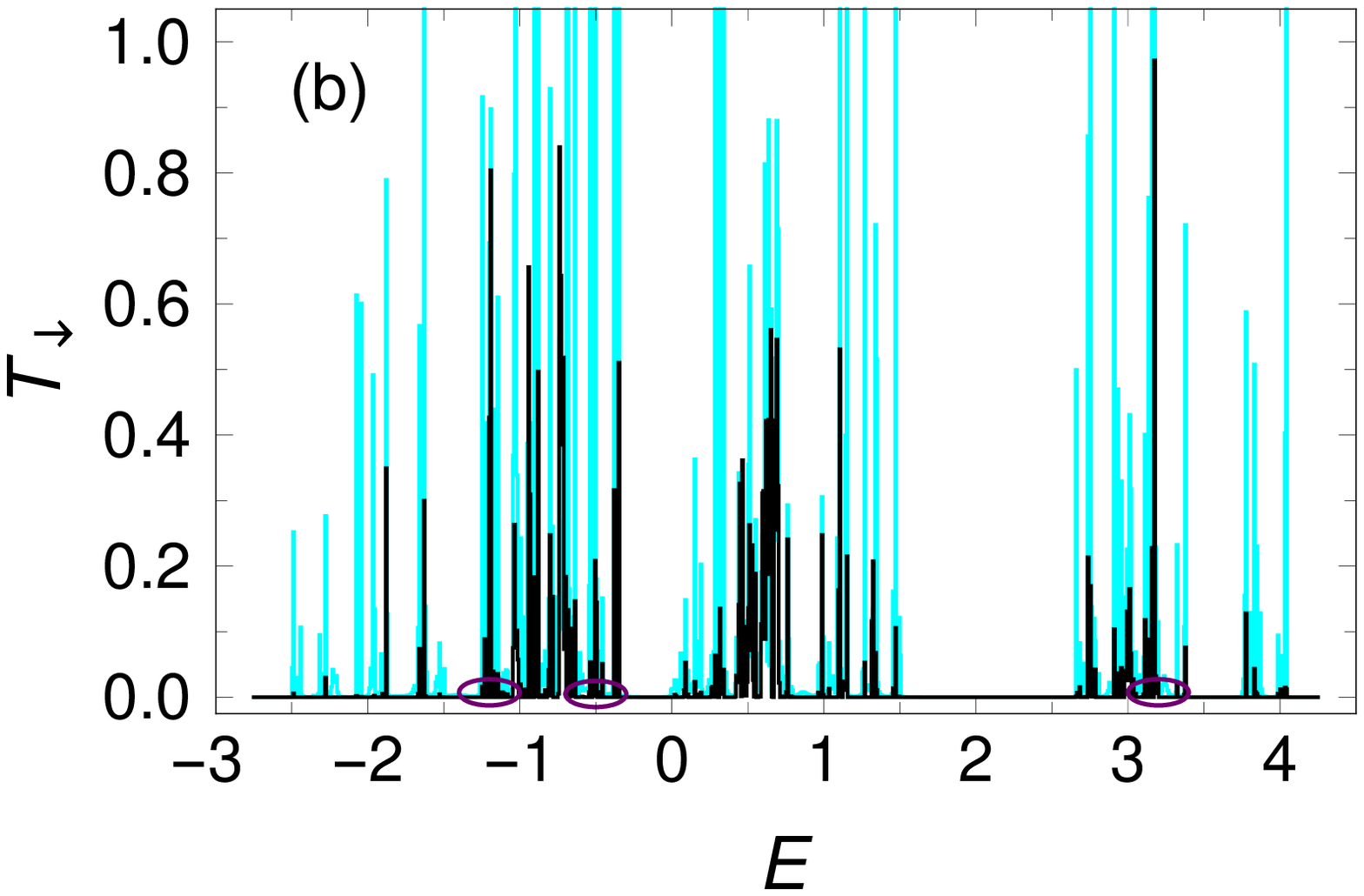}
\caption{(Color online). Behavior of (a) spin-up (red color) and (b) spin-down (black color) transmission probabilities as a function of energy $E$ superimposed on the total density of states (cyan color) for an 8th generation SPG network, with noncollinear spin configuration. The dark magenta ellipses denote the existence of spin-dependent mobility edges.}
\label{gen8}
\end{figure*}
\subsubsection{Spin-resolved bond current densities}
To visualize the distribution of the spin-resolved currents at each bond, we compute the spin-resolved bond current densities from site $j$ to $i$. The bond current density effectively describes the flow of charges, while the spin-dependent bond current density $J_{i\sigma\leftarrow j\sigma^\prime}$ illustrates the flow of spins, which starts at site $j$ with spin $\sigma^\prime$ and end up at site $i$ with spin $\sigma$, with the spin-quantization axis along the $z$-direction. The spin-dependent bond current density can be evaluated with the following expression~\cite{bcd,bcd1}
\begin{equation}
J_{i\sigma\leftarrow j\sigma^\prime} = \frac{2e}{\hslash}\text{Im}\left[ \left(\psi_i^\sigma\right)^* H_{ij} \psi_j^{\sigma^\prime} \right] = \frac{2e}{h}\text{Im}\left[ H_{i\sigma, j\sigma^\prime} {\mathcal G}_{i\sigma, j\sigma^\prime}^n \right],
\label{eq:bc}
\end{equation}
where $H_{i\sigma, j\sigma^\prime}$ is the element of the Hamiltonian matrix $\left(H_{\text{SPG}}\right)$ corresponding to the $\sigma$ spin at site $i$ and $\sigma^\prime$ spin at site $j$. $\psi_i^{\sigma}$ is the amplitude of the electronic wave function with spin $\sigma$ at site $i$. ${\mathcal G}_{i\sigma, j\sigma^\prime}^n$ denotes the matrix element of the correlation function corresponding to the $\sigma$ spin at site $i$ and $\sigma^\prime$ spin at site $j$. The correlation function ${\mathcal G}^n$ is defined as 

\begin{equation}
{\mathcal G}^n = {\mathcal G}^r \Gamma_S {\mathcal G}^a.
\end{equation} 
The correlation function is computed by setting the occupation probability of the source to unity and that of the drain to zero.

Finally, we define the net up and down spin bond current densities as
\begin{eqnarray}
J^\uparrow_{ij} = J_{i\uparrow\leftarrow j\uparrow} + J_{i\uparrow\leftarrow j\downarrow},\label{bc1}\\
J^\downarrow_{ij} = J_{i\downarrow\leftarrow j\uparrow} + J_{i\downarrow\leftarrow j\downarrow},\label{bc2}
\end{eqnarray}
where $J^\uparrow_{ij}$ denotes the net up spin bond current density and $J^\downarrow_{ij}$ refers to the net down spin bond current density from site $j$ to site $i$.


\section{\label{sec:results}Numerical Results and Discussions}
Before we begin, let us first mention the parameter values used in the present work. All the energies are measured here in the unit of eV. The on-site energies in the fractal network as well as in the source and drain electrodes are set at zero. The NNH strength for the SPG is considered as $t=1$, while that for the leads are taken as $t_0=2.5$, to work within the wide-band limit.  The coupling strengths of the SPG network to the source and drain electrodes, characterized by the parameters $\tau_S$ and $\tau_D$, are also fixed at $1$. For any other set of parameter values, the physical picture will be qualitatively the same, which we confirm through our detailed numerics. Unless stated, the strength of the spin-dependent scattering factor is fixed at $h=0.5\,$eV, and the spin-resolved transmission coefficients, currents, and polarizations are computed for the spin-quantization direction along the $z$-axis.

\subsection{Spin-resolved transmission coefficients and currents, polarization: Absence of light}
We start our discussion by analyzing the spin-resolved two-terminal transmission coefficients of an SPG network in the absence of light. In Figs.~\ref{gen8}(a) and (b), we show the spin-resolved transmission coefficients as a function of energy. The spin-up transmission probability $T_\uparrow$ is denoted with red color and the spin-down transmission probability $T_\downarrow$ by black color. As the Hamiltonian of our system cannot be decoupled for the up and down spin electrons, we compute the total density of states (DOS) and superimposed it (denoted with cyan) on the spin-resolved transmission probabilities to detect the mobility edge, if there is any. Here it is important to note that all the states of an SPG lattice become localized in the asymptotic limit due to the structure-induced localization~\cite{kallol-scirep}. Therefore, whether the localization behavior persists in the asymptotic limit for the noncolinear AFM SPG, we consider a bigger SPG (8th generation) that contains a fairly large number of lattice sites. Unlike the spin-less case, the noncolinear spin arrangement potentially transforms the completely localized SPG network into a partially conducting one which can be seen from the large values of the transmission probabilities at certain energy values. The most important feature in Figs. \ref{gen8}(a) and (b) is that the behavior of the spin-up and spin-down transmission probabilities are different from each other. This ensures a finite spin polarization for the $z$-component which is completely due to the broken spin sublattice symmetry.

We also detect multiple spin-dependent mobility edges in Figs.~\ref{gen8}(a) and (b). For instance, in Fig.~\ref{gen8}(a), near the energy $E \sim -1.25$, there is a fine strip of non-zero transmission coefficient for the up spin electrons, which is a manifestation of extended states. Again, to the immediate left/right of the strip, though the DOS is finite, the vanishing spin-up transmission coefficient indicates that the states are localized. This is a typical example of a mobility edge associated with the up spin electrons. Similar features are also observed in Fig.~\ref{gen8}(b) where we detect multiple spin-down mobility edges. The region across a mobility edge is marked with a dark magenta ellipse in Figs.~\ref{gen8}(a) and (b) for better visualization. The existence of spin-dependent mobility edges certainly makes the present work more technologically intriguing, where {\it a noncolinear AFM SPG may be utilized as a spin-based switching device}.

Based on the transmission spectra as discussed in Figs.~\ref{gen8}, let us concentrate on the characteristic features of spin-resolved currents and polarization, which is the central focus of the present work. 
\begin{figure}[ht!]
\centering
\includegraphics[width=0.2325\textwidth,height=0.155\textwidth]{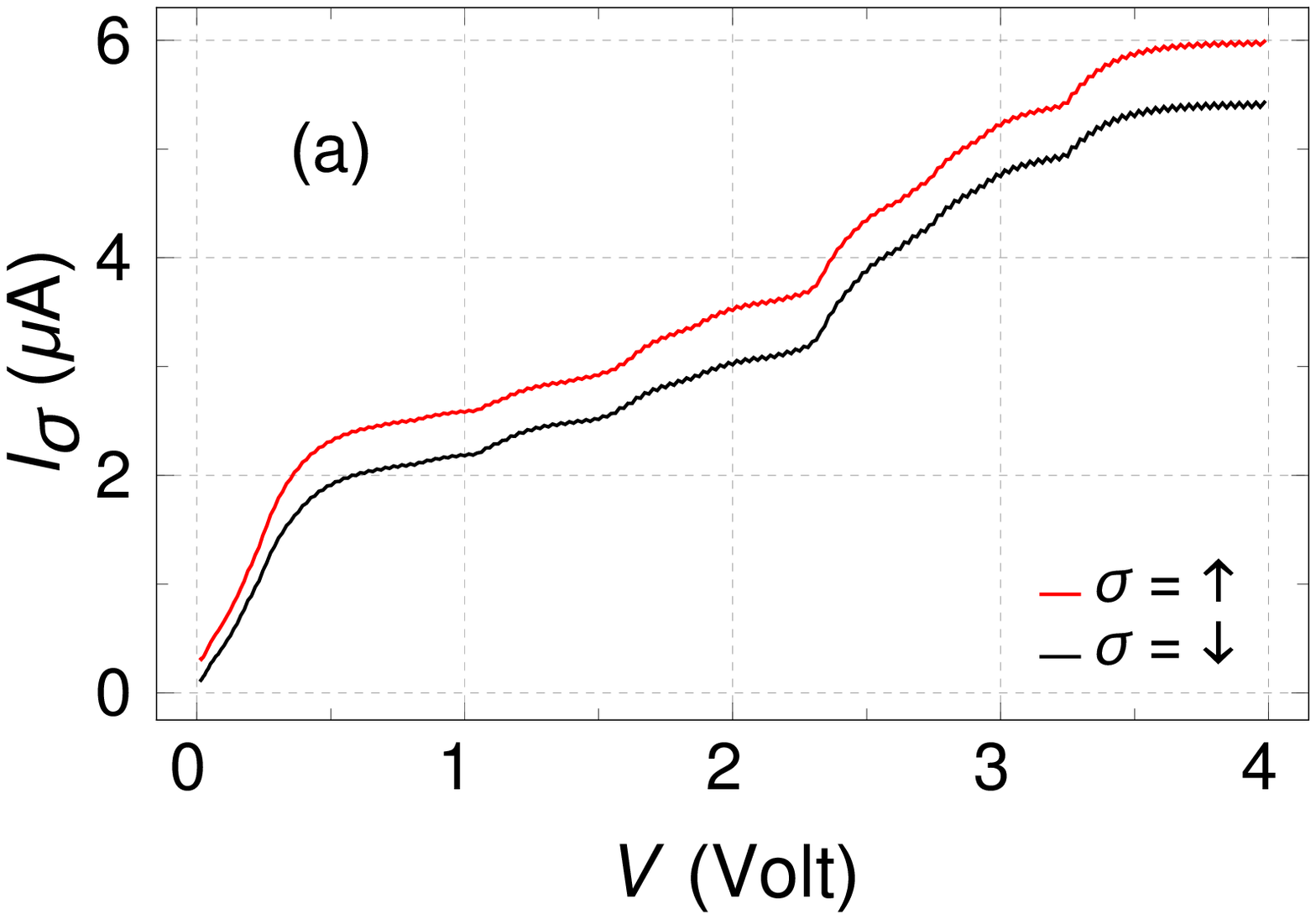} \hspace{0.1cm}
\includegraphics[width=0.2325\textwidth,height=0.155\textwidth]{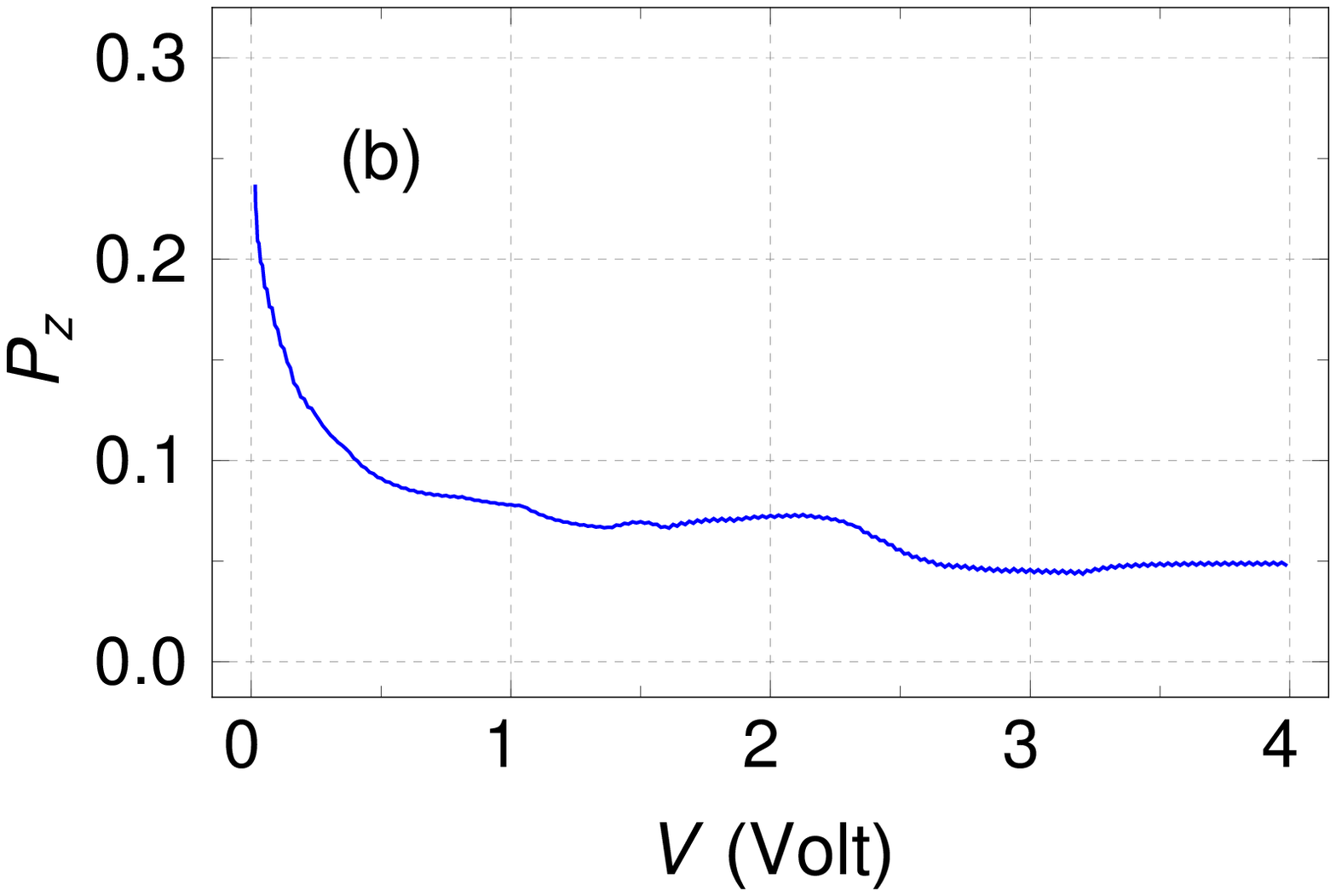}
\caption{(Color online). (a) Spin-resolved currents and  (b) spin polarization coefficient $P_z$ as a function of bias voltage in the absence of light. Spin-up and spin-down currents are denoted with red and black colors respectively. The Fermi energy is fixed at $E_F=0.5\,$eV.}
\label{IPgen8}
\end{figure}
Figures.~\ref{IPgen8}(a) and (b) represent the behavior of spin-resolved currents and polarization as a function of the bias voltage. The spin-resolved currents are computed using Eq.~\ref{def-curr}. The Fermi energy is fixed at $E_F = 0.5\,$eV.
\begin{figure*}[t!]
\centering
\includegraphics[width=0.45\textwidth]{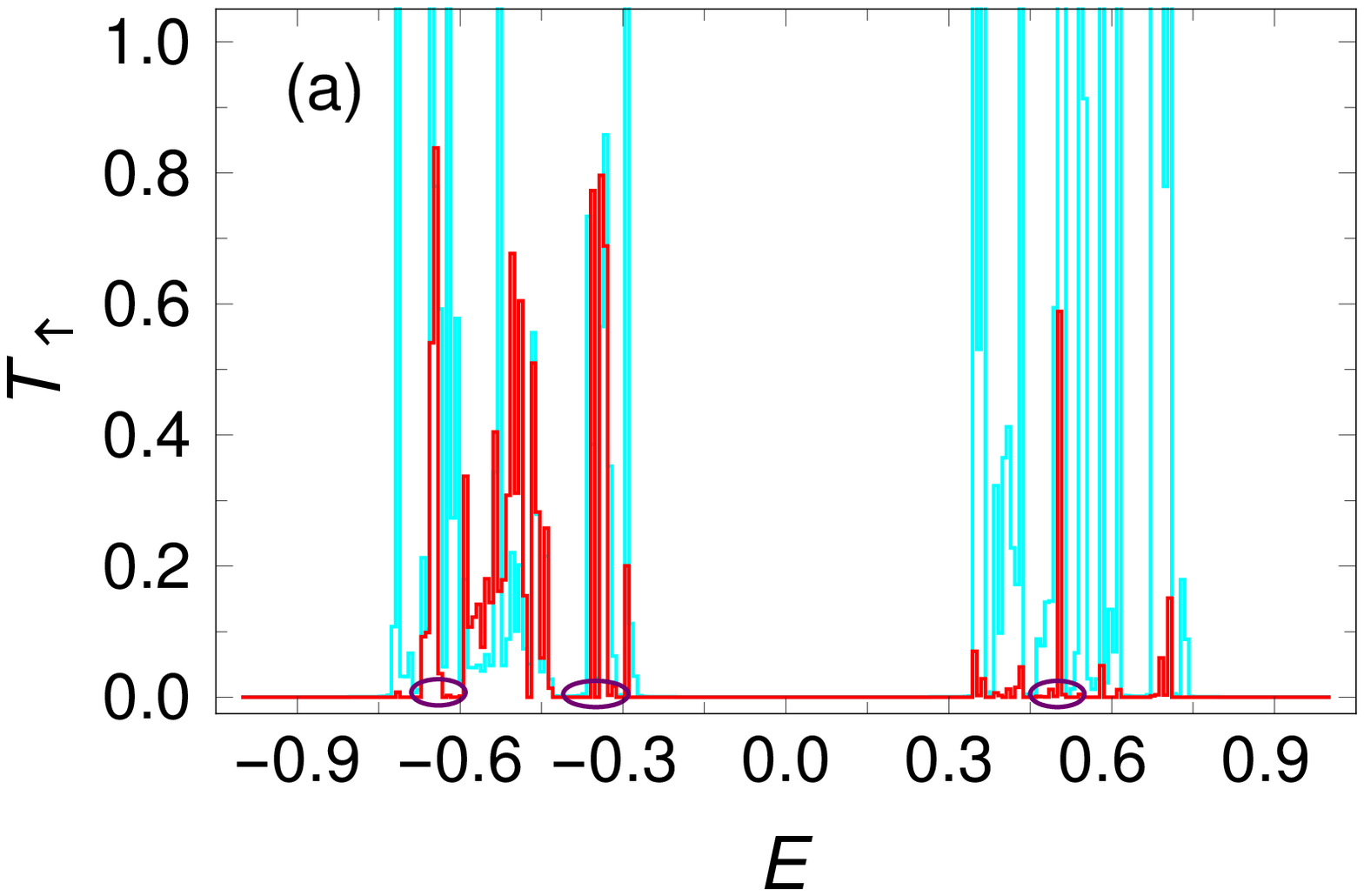} \hspace{0.5cm}
\includegraphics[width=0.45\textwidth]{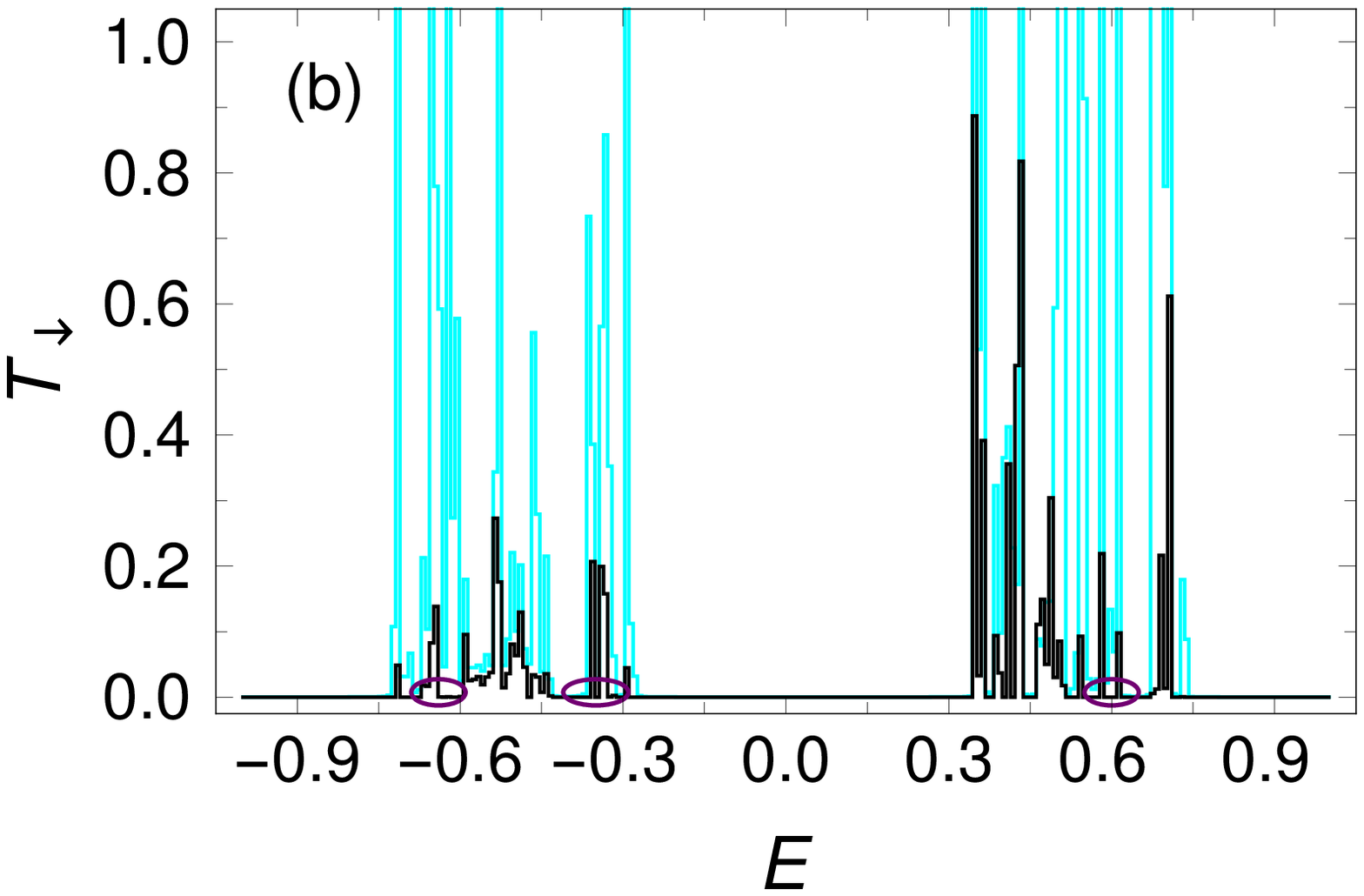}
\caption{(Color online). Behavior of (a) spin-up (red color) and (b) spin-down (black color) transmission probabilities as a function of energy $E$ superimposed on the total density of states (cyan color) in the presence of light. The light parameters are $A_x =2.5, A_y = 2$, and $\phi = \pi/2$. The dark magenta ellipses denote the existence of spin-dependent mobility edges. The generation of SPG and all other physical parameters are similar to Fig.~\ref{gen8}.}
\label{gen8wl}
\end{figure*}

In Fig.~\ref{IPgen8}(a), the spin-up and spin-down currents are denoted with red and black colors respectively. The spin currents are of the order of $\mu$A and they increase with the bias voltage. The increasing behavior of the spin currents is obvious from Eq.~\ref{def-curr}. Increasing the bias means increasing the allowed energy window, that is, more transmission peaks appear within the bias window. Consequently, the current increases with the bias voltage. As expected from the spin-resolved transmission spectra, the spin currents due to the up and down spin electrons differ from each other throughout the voltage window. This results in a non-zero spin polarization, as shown in Fig.~\ref{IPgen8}(b). The noted maximum polarization is about $25\%$ for very low bias voltage, and the degree of polarization decreases further as the voltage increases. It is quite a significant result in the sense that despite any spin-splitting interaction like spin-orbit coupling, a noticeable degree of spin polarization is achieved just by choosing a specific spin configuration. Such a spin orientation in kagome lattice are dubbed as $\mathbf{Q} =0$ configuration ($Q$ is known as magnetic wave vector), are already explored extensively~\cite{km1,km2,Ohgushi,Chen-PRL} and appear in many realistic materials even at room temperature~\cite{Nayak,Nakatsuji,Zhang-JPCM,Kubler,she2,Zhang-PRB-2017}. In summary, we can say that these types of spin configurations induce the same spin-splitting effect by breaking the spin rotational symmetry, analogous to the spin-orbit coupling. However, though we achieve a spin-splitting effect in the absence of SOC, the charge-to-spin conversion ratio is not up to the mark. In the next section, we provide a new prescription to enhance the spin polarization significantly by irradiating the system.


\subsection{Spin-resolved transmission coefficients, spin-dependent currents, and spin polarization: Presence of light}
We begin our discussion by analyzing the spin-resolved two-terminal transmission coefficients of an SPG network in the presence of light. In Figs.~\ref{gen8wl}(a) and(b), we show the up and down spin transmission coefficients, superimposed with the total DOS in the presence of light for an 8th generation SPG AFM network. The light parameters are $A_x =2.5, A_y = 2$, and $\phi = \pi/2$. Similar to the results obtained earlier in the absence of light, the spin-resolved transmission spectra are also associated with multiple mobility edges. But, the introduction of light irradiation makes the mobility edges more prominent. Here too, the region across a mobility edge is marked by a dark magenta ellipse in Fig.~\ref{gen8wl}(a) and (b) for better viewing. 

The interesting feature observed from Figs.~\ref{gen8wl}(a) and (b) is the behavior of spin-resolved transmission probabilities $T_\uparrow$ (red color) and $T_\downarrow$ (black color), which are completely different from each other. To be more specific,  the $T_\uparrow$-$E$ spectrum is divided into two branches associated with a gap about the zero energy. $T_\uparrow$ has more transmission values on the left side of the zero-energy than the right side. On the other hand, the behavior of the down spin transmission spectrum is opposite to that of the up spin transmission spectrum. This is an ideal situation to achieve a high degree of polarization, where the Fermi energy can be placed in such a way that at the fixed $E_F$, one specific spin band gets suppressed while the opposite band shows higher transmission values. When the SPG is irradiated with light, the effective hopping gets renormalized following the relation given in Eq.~\ref{effhop}. Moreover, as the modification of the hopping integrals depends on the bond directions, a spatial anisotropy is established in the SPG network. As a result of that, and also due to the noncolinear magnetic texture, the spin channels are greatly modified but differently, which explains the significant change in the behavior of the spin-resolved transmission spectra. We also note that the allowed energy window gets shortened due to the modified hopping integrals in the presence of irradiation.

\begin{figure}[ht!]
\includegraphics[width=0.235\textwidth,height=0.155\textwidth]{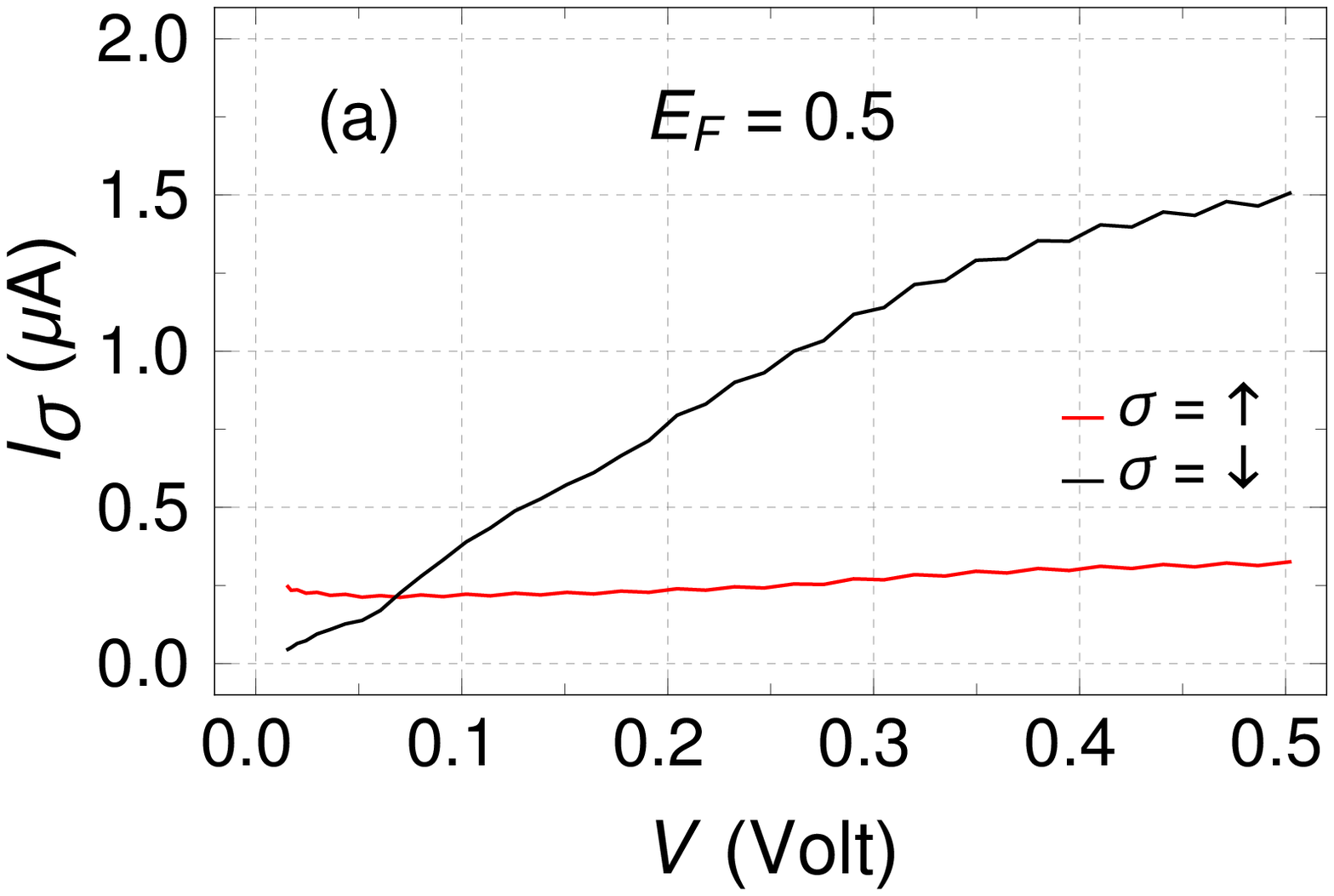} \hfill
\includegraphics[width=0.235\textwidth,height=0.155\textwidth]{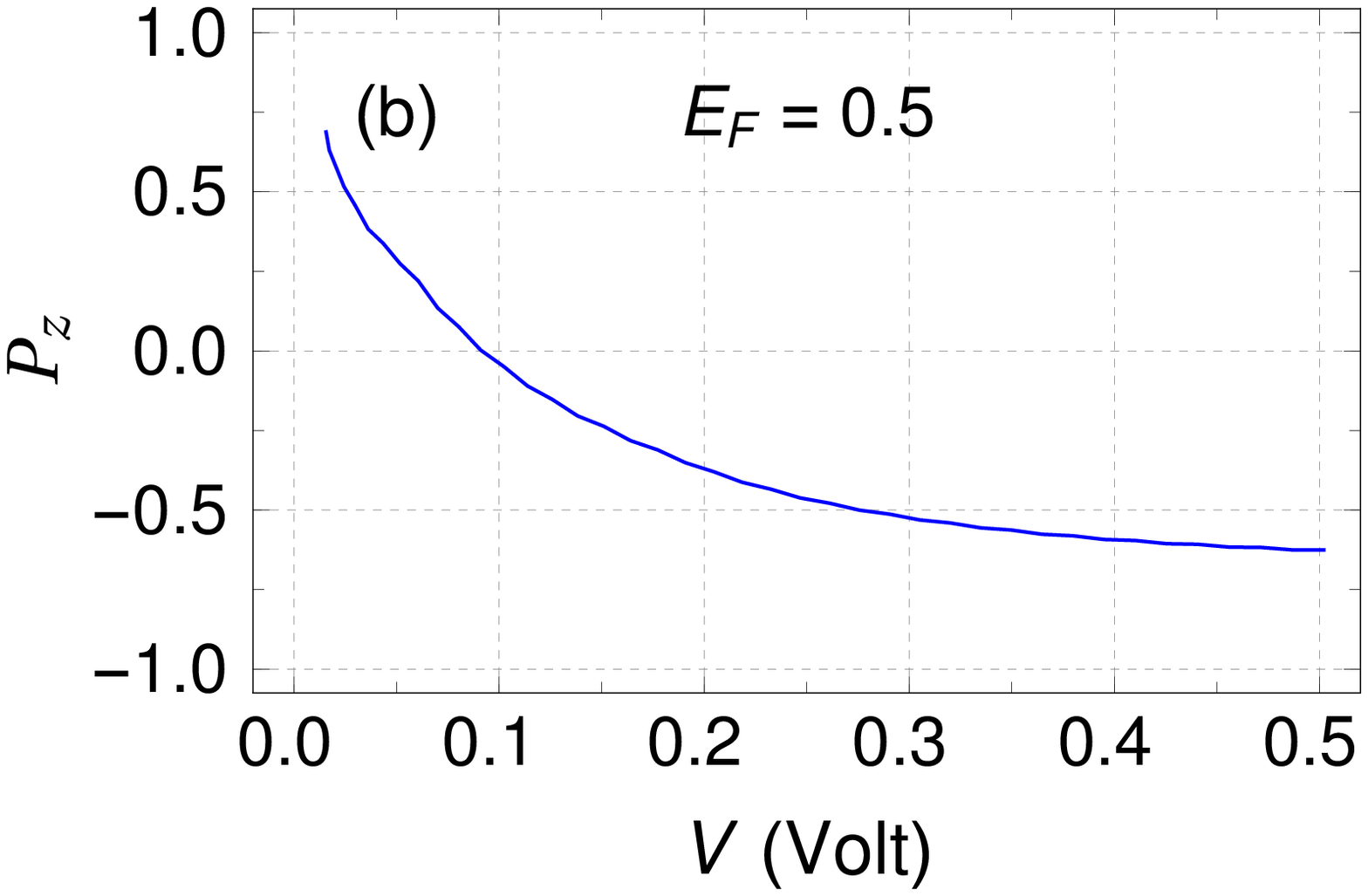}\vskip 0.1 in
\includegraphics[width=0.235\textwidth,height=0.155\textwidth]{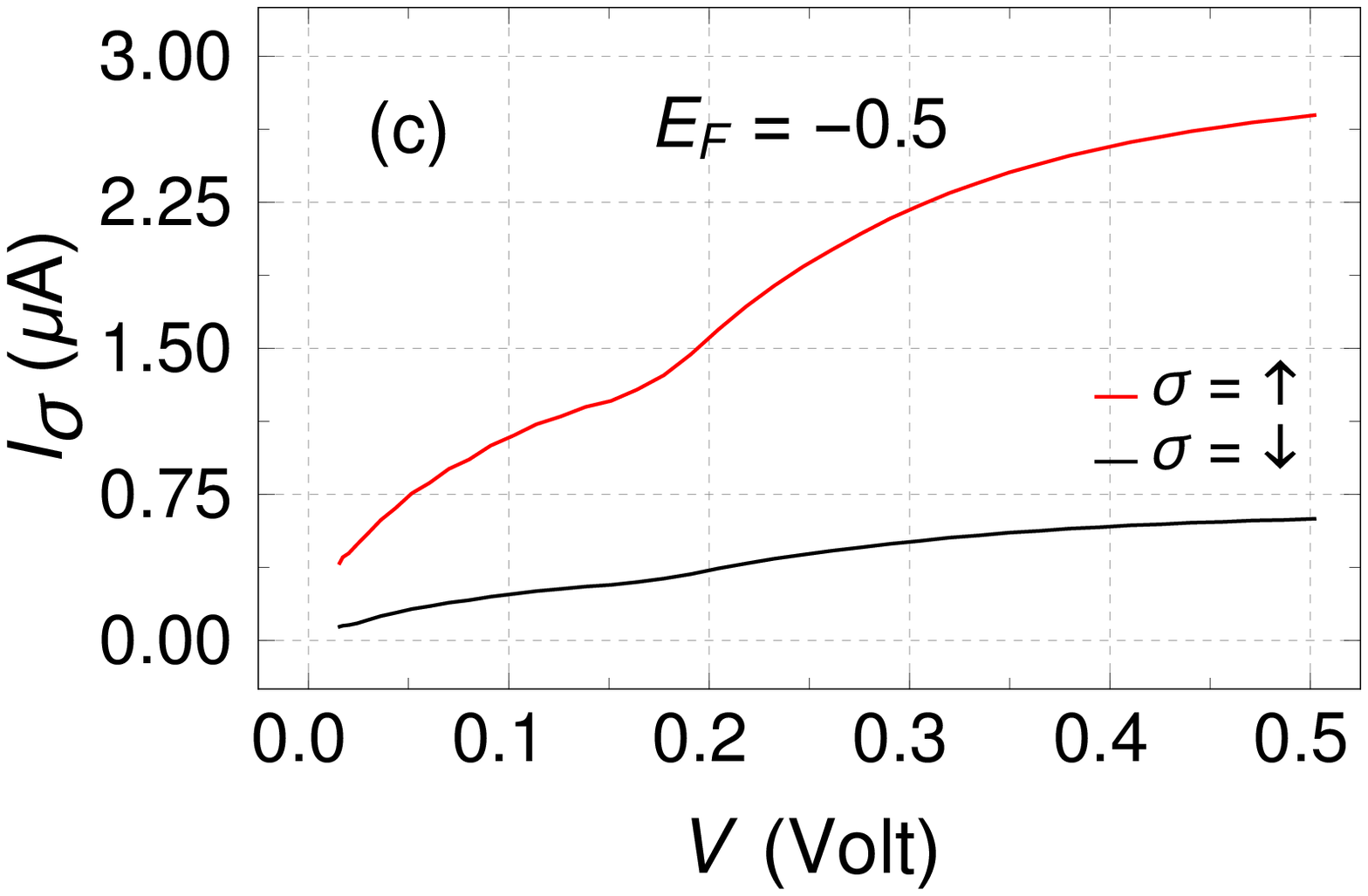} \hfill
\includegraphics[width=0.235\textwidth,height=0.155\textwidth]{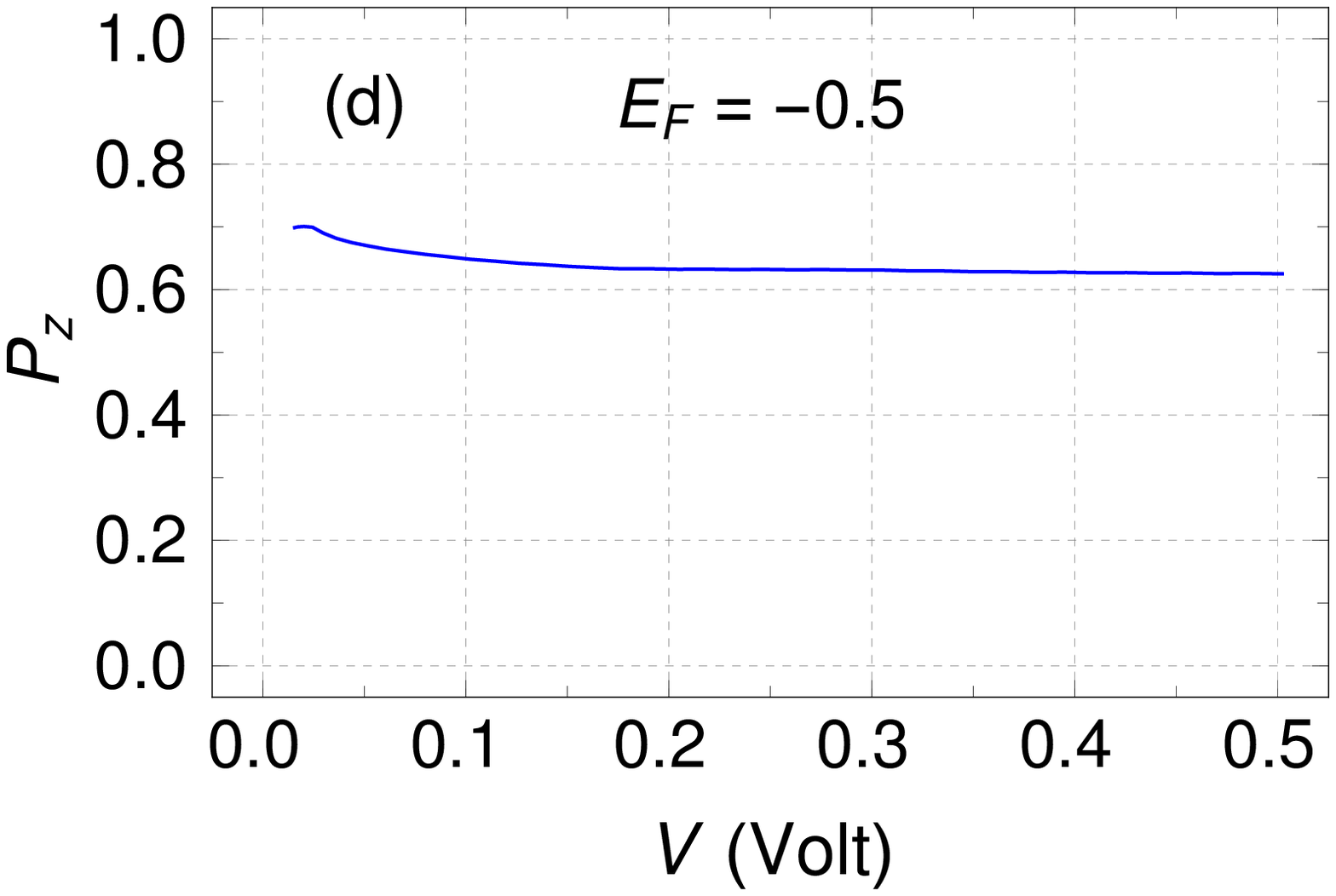}
\caption{(Color online). (a) Spin-resolved currents and  (b) spin polarization coefficient $P_z$ as a function of bias voltage in the presence of light for $E_F=0.5$. (c) Spin-resolved currents and  (d) spin polarization coefficient $P_z$ as a function of bias voltage for $E_F=-0.5$. The light parameters are $A_x =2.5, A_y = 2$, and $\phi = \pi/2$. All other the physical parameters and color conventions are identical with Fig.~\ref{IPgen8}.}
\label{IPgen8wl}
\end{figure}

With the knowledge of the electronic transmission profile, it is now easier to explain the spin-resolved current-voltage characteristics and spin polarization of the SPG network in the presence of light. The results are presented in Fig.~\ref{IPgen8wl}, where the variations of spin-dependent currents and the spin polarization coefficient are shown as a function of bias voltage. The choice of Fermi energy is always important, as the degree of spin polarization and its sign can be manipulated by setting the Fermi energy at appropriate places within the allowed energy window. Therefore, in Figs.~\ref{IPgen8wl}(a), (b) 
and Figs.~\ref{IPgen8wl}(c), (d), we set the Fermi energies at $E_F =0.5$ and $-0.5$ respectively. In Fig.~\ref{IPgen8wl}(a), we see that for smaller bias voltage, the up-spin current is higher than the down-spin current. After that, the down spin current becomes more dominant. Consequently, the polarization changes signs from positive to negative as is seen from Fig.~\ref{IPgen8wl}(b). The corresponding maximum polarization is obtained around 60\%. In Fig.~\ref{IPgen8wl}(c), we see that the up spin current is always higher than the down spin current throughout the voltage window, and therefore, the polarization is positive. In this case, the maximum polarization is found to be more than 60\%, as shown in Fig.~\ref{IPgen8wl}(d). Overall, the irradiation enables us to achieve a moderate spin polarization even for a large SPG (8th generation) and with appropriate choices of the light parameters, it is also possible to get a very high degree of spin polarization, which we shall also explore in the present work.

\subsubsection{Role of spin-dependent scattering factor $h$}
\begin{figure}[ht!]
\centering
\includegraphics[width=0.45\textwidth]{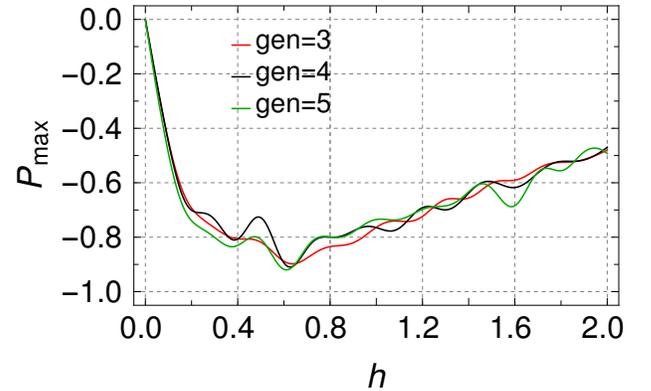} 
\caption{(Color online). Maximum polarization $P_{\rm max}$ with sign as a function spin-dependent scattering parameter $h$ for three different generations. The light parameters are $A_x =2.5, A_y = 2$, and $\phi = \pi/2$. The Fermi energy is fixed at $E_F =0.5$. The red, black, and green colors are corresponding to the results for generations 3, 4, and 5, respectively.}
\label{PmaxVSh}
\end{figure}

The presence of spin-dependent scattering interaction plays an important role in spin-dependent transport phenomena. In order to understand the behavior of spin polarization with the strength of spin-dependent scattering factor $h$, we have presented the maximum of polarization $P_{\rm max}$ as a function of $h$ for a particular set of light parameters for three different generations as shown in Fig.~\ref{PmaxVSh}. We compute $P_z$ by varying the
bias in the allowed voltage window keeping the Fermi energy fixed at $E_F=0.5$ and then take the maximum of $P_z$ with sign, which we refer to as $P_{\rm max}$.  The results for generations 3, 4, and 5 are denoted with red, black, and green colors respectively. The light parameters are considered here as $A_x=2.5, A_y = 2 $, and $\phi= \pi/2$. The overall envelope of the $P_{\rm max}$-$h$ curve for the three different generations is more or less similar. The spin-dependent scattering strength is varied within the window $0$ to $2$. We see that within the given window of $h$, the maximum polarization is about $90\%$ for $h\sim 0.5$.  Here it should be noted that the magnitude of $h$ can be higher than the that considered in the present work due to strong coupling between the itinerant electrons and the moments~\cite{Su-APL}. This is one of the key advantages of the strong spin-dependent scattering in a magnetic material compared to the spin-orbit-coupled systems. As within the given $h$-window, the maximum polarization is observed for $h\sim 0.5$, we fix the strength of the spin-dependent scattering factor as $h=0.$ in the rest of the work.


\subsubsection{Explicit dependence of light parameters on spin polarization}
\begin{figure}[ht!]
\includegraphics[width=0.235\textwidth,height=0.23\textwidth]{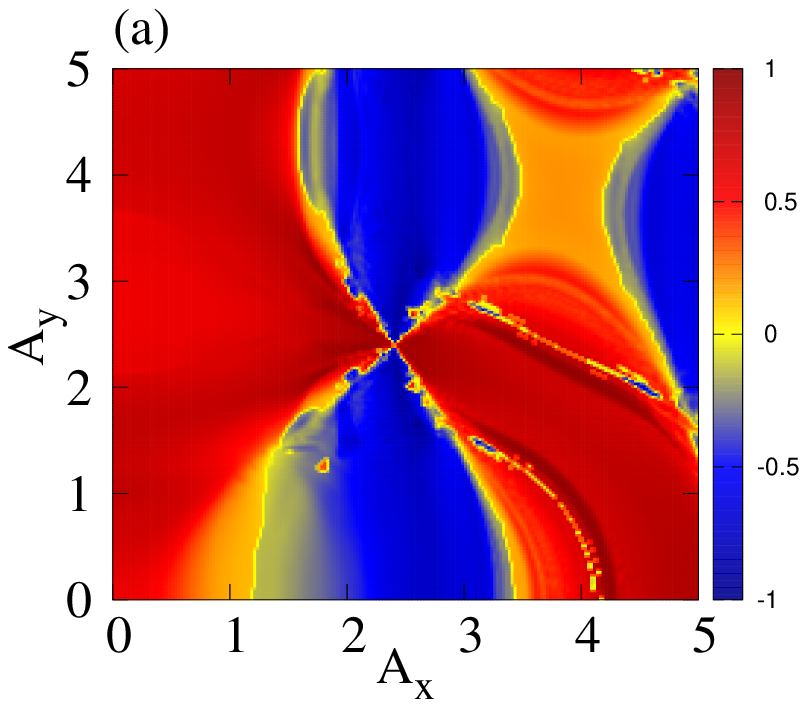} \hfill
\includegraphics[width=0.235\textwidth,height=0.23\textwidth]{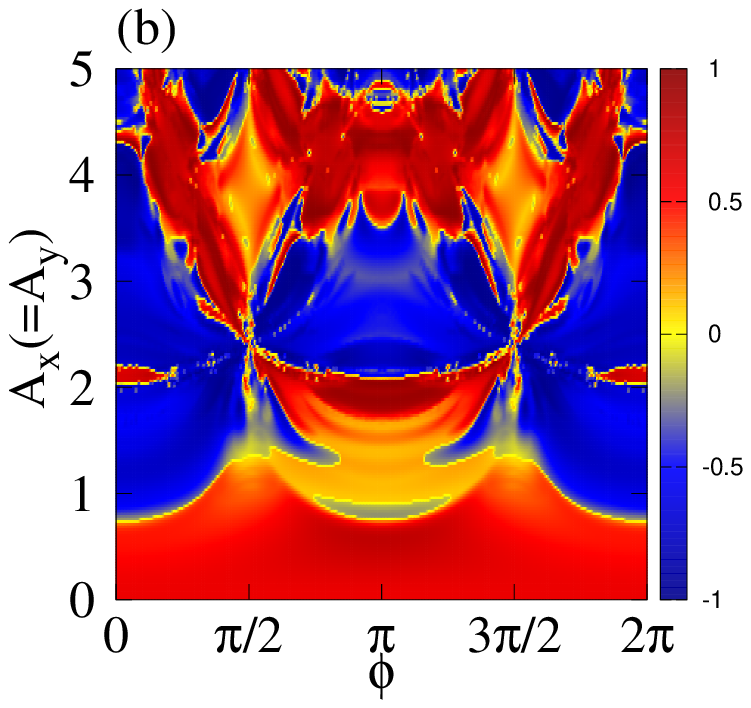} 
\caption{(Color online). Color density plot for $P_{\rm max}$ as functions of (a) $A_x$ and $A_y$ with $\phi = \pi/2$ (b) $A_x\left(=A_y\right)$ and $\phi$ for a 4th generation  SPG network. Dark red and dark blue regions denote positive and negative high degree of spin polarizations, respectively.}
\label{density}
\end{figure}

To illustrate the explicit dependence of light on spin polarization, here, we explore the effects of all the light parameters $A_x, A_y$, and $\phi$ by varying them over a wide range in the parameter space. Under this situation, we can potentially investigate the effects of all kinds of polarized lights, viz., linear, elliptical, circular. Figure~\ref{density}(a) shows the density plot of $P_{\rm max}$ as functions of $A_x$ and $A_y$, keeping $\phi$ fixed at $\pi/2$ for a 4th generation SPG network. The spin-dependent scattering parameter is set at $h=0.5$. The definition of $P_{\rm max}$ is the same as mentioned earlier in the discussion of Fig.~\ref{PmaxVSh}.  The field amplitudes $A_x$ and $A_y$ are varied from 0 to 5 to examine the spin polarization. The dark red and dark
blue regions correspond to the positive and negative high degrees of spin polarization, respectively. The density plot (Fig.~\ref{density}(a)) reveals that significant polarization can be achieved for a wide range of light parameters, and the maximum polarization can be more than $90\%$. Moreover, the sign of the spin polarization can be tuned with an appropriate set of light parameters. Figure~\ref{density}(b) shows the density plot of $P_{\rm max}$ as functions of $A_x(=A_y)$ and the phase factor $\phi$ for a 4th generation SPG network. A nice pattern is emerged in Fig.~\ref{density}(b). Large spin polarization is observed for both up and down spin electrons. Moreover, $P_\text{max}$ shows a symmetric nature around the $\phi=\pi$ line. This symmetric nature of $P_\text{max}$ is described as follows. The SPG network has three different hopping terms, One in the horizontal direction, and the other two, along the angular directions (see Fig.~\ref{spg}). In the presence of light, the hopping terms get renormalized and they are directional dependent by Eq.~\ref{effhop}. We find that the horizontal hopping term is independent of phase $\phi$, while the two angular hopping terms are not. For the transformation, $\phi \rightarrow \phi + \pi$, the status of the two angular hopping terms get swapped. But, the Hamiltonian remains the same under this transformation. As a result, the $P_{\rm max}$ becomes symmetric about $\phi=\pi$.

What we gather so far is that the presence of irradiation makes the mobility edges more prominent. The splitting between up and down spin current enhances as the system is exposed to irradiation, leading to a significant enhancement in spin polarization. Thus, we can engineer spintronic devices using noncolinear AFM SPG, where the spin polarization can be tuned externally with the help of light parameters.
This phenomenon undoubtedly yields a new signature of controlling spin selective electron transfer.

\subsubsection{Spatial distribution of spin-dependent bond current density}

\begin{figure}[ht!]
\centering
\includegraphics[width=0.22\textwidth]{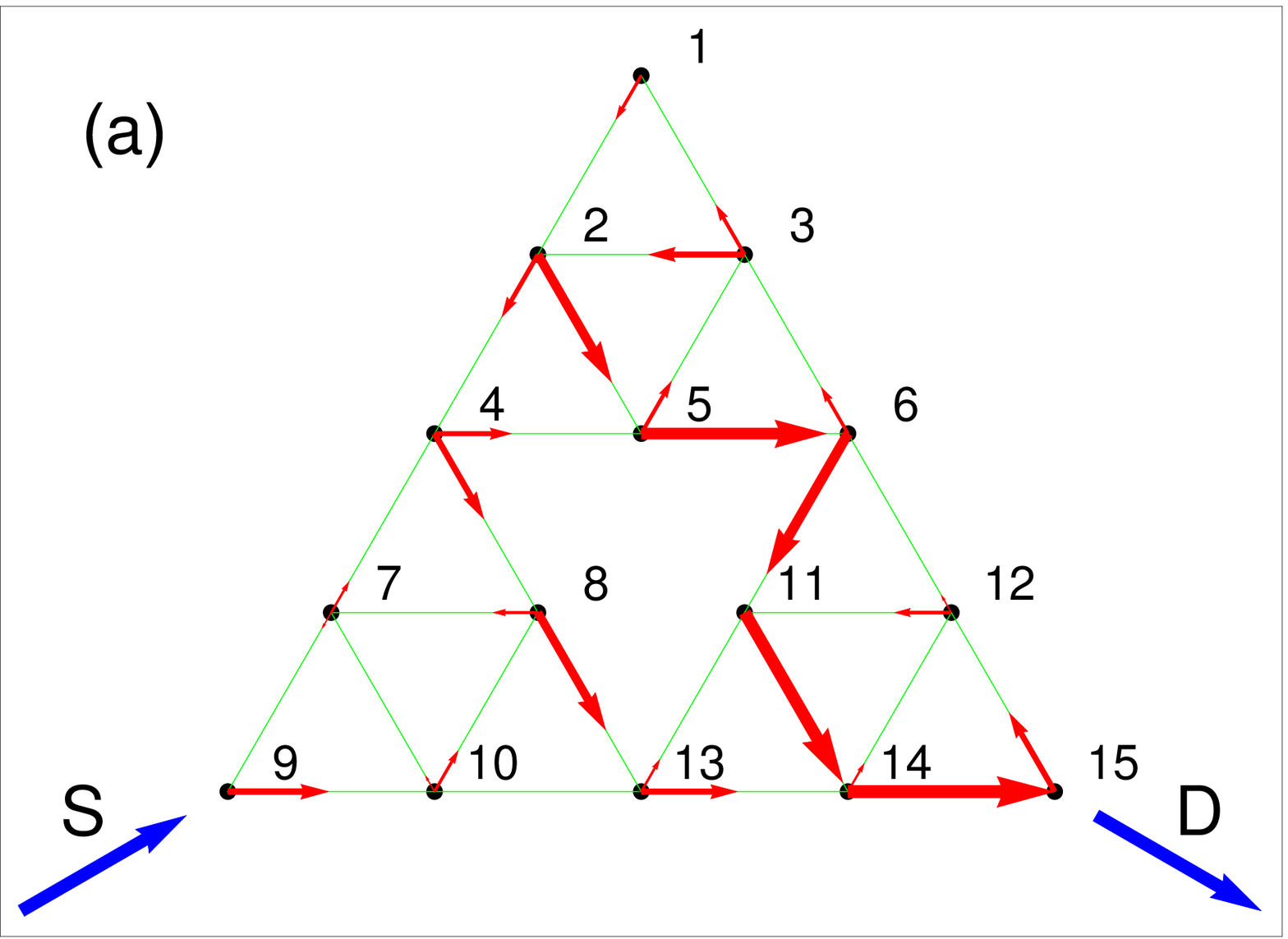} \hspace{0.25cm}
\includegraphics[width=0.22\textwidth]{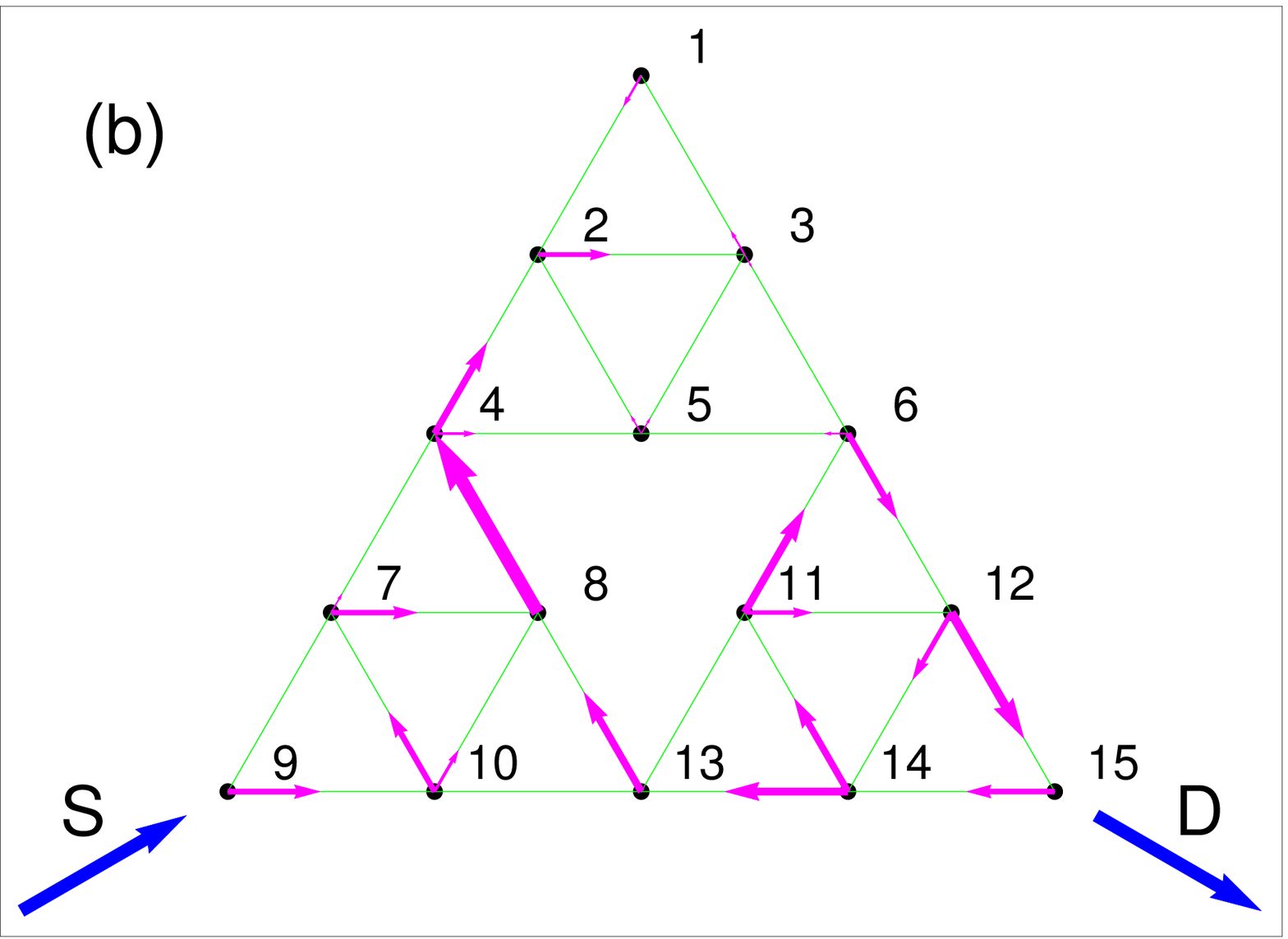} \hspace{0.25cm}
\caption{(Color online). The spatial distribution of (a) spin-up (denoted with red color) and (b) spin-down (denoted with magenta color) bond current densities  of a 3rd generations SPG network in presence of light. The light parameters are $A_x = 0.75$, $A_y =1.1$, and $\phi=\pi/4$. The energy is fixed at $E=-0.1$. The length of the arrows and the size of the arrowheads indicate the magnitude of the bond current density at each bond, while the directions of the currents are shown by the arrowheads. The absence of arrow in the bonds implies either the bond current density is zero at the specific bond or the magnitude is vanishingly small. The number at the vicinity of each lattice point in the schematic diagram indicates the site index.}
\label{BCgen4}
\end{figure}

To investigate the microscopic profile of local spin current densities in a finite-size SPG network i.e., how the current densities for the up and down spin electrons are distributed among the different bonds of the SPG network, we compute the spin-dependent bond current densities in the presence of light. The spatial distributions of the bond current density for the up and down spin electrons are depicted in Fig.~\ref{BCgen4}(a) and (b), respectively. For better visualization, we consider a 3rd generation SPG network in the present case. The thick blue arrows in the left and right of each diagram denote the positioning of the source and drain, respectively. 
The bond current densities are calculated following the Eqs.~\ref{bc1} and \ref{bc2}, where we consider the spin-quantization axis along the $z$-direction. We consider light parameters as $A_x = 0.75$, $A_y =1.1$, and $\phi=\pi/4$. The Fermi energy is set at $E =-0.1$. The bond current distribution for the up and down spins are denoted with red and magenta colors, respectively. The length of the arrows and the size of the arrowheads indicate the magnitude of the bond current density at each bond. The directions of the currents are shown by the arrowheads. The absence of an arrow in the bonds implies either the bond current density is zero at the specific bond or the magnitude is vanishingly small.

The first impression that is obtained from  Fg.~\ref{BCgen4} is that the spatial distribution of spin-up bond current density is distinctly different from its down counterpart. A careful inspection shows that for some specific bonds, the current density is vanishingly small, while for some other bonds, they are large. This particular feature is observed for both spin-up and spin-down current densities. This can be understood from the fact that the presence of light irradiation renormalizes the hopping integral, which are directional dependent. On the other hand, there is a noncolinear magnetic texture, which induces a spin-dependent scattering phenomenon. Due to the combined effect of these two, the parity between the up and down spin currents is lost, yielding an effective spin polarization.

\subsubsection{Interface Sensitivity}
\begin{figure}[ht!]
\centering
\includegraphics[width=0.45\textwidth]{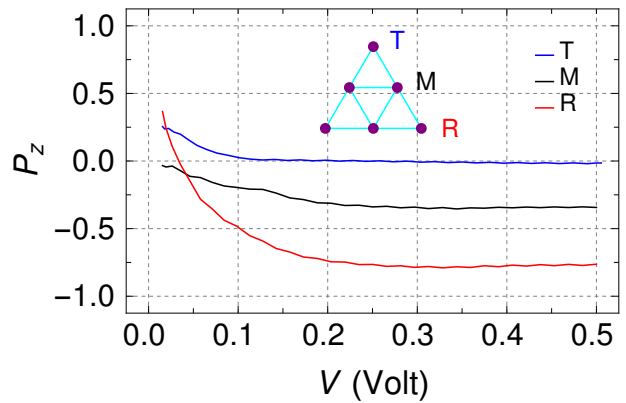} 
\caption{(Color online). Spin polarization coefficient $P_z$ as a function of bias voltage for three different drain positions for a 5th generation SPG network. Three different lead positions are considered, namely, top vertex (denoted with blue color), bottom right vertex (denoted with red color), and at the middle position between these two vertices (denoted with black color) of the SPG triangle, as shown by the inset for illustration.  The chosen  light parameters are $A_x =2.5, A_y = 2 $ and  $\phi =\pi/2$ and the Fermi energy is set at $E_F =0.5$.}
\label{interface}
\end{figure}

The quantum interference among the electronic waves passing through different branches of the SPG significantly modifies the transport properties and the modification becomes more effective in multi-loop geometries. Therefore, we need to study the degree of spin polarization for different lead positions attached to the SPG network, as the search for a favorable interface geometry is extremely important. In Fig.~\ref{interface}, we present the spin polarization as a function of bias voltage for three different drain positions. A 5th generation SPG network is considered here. We particularly choose three different drain positions, namely the top vertex (denoted with blue color), 
bottom right vertex (denoted with red color), and at the middle position between these two vertices (denoted with black color) of the SPG triangle, as shown by the inset in Fig.~\ref{interface}. The light parameters are considered here as $A_x= 2.5, A_y = 2$, and $\phi= \pi/2$. The Fermi energy is fixed at $E_F =0.5$. We see that when the drain is connected at the bottom right vertex, $P_z$ shows higher values throughout the bias window than the other two drain positions. For the right vertex, the spin polarization coefficient is about $0.8$, dominated by the down spin electrons. When the drain is connected at the middle between the top and right vertices, $P_z$ is about $0.25$. For the top vertex case, the degree of spin polarization is very poor. For lower bias, $P_z$ is about 0.1, and then with increasing the bias voltage, $P_z$ decreases to zero. It turns out that for other choices of light parameters, the features of the interface sensitivity remain the same.

Using symmetry argument, one may argue that as the drain position for the top and the right vertices are equivalent with respect to the source, should result in the same spin polarization. Though geometrically, this seems correct, as soon as the system is exposed to irradiation, the symmetry argument is no longer valid. The presence of light modifies the hopping integrals according to Eq.~\ref{effhop}, incorporating asymmetry in the hopping integral in the angular and the horizontal directions. As a result, the spin polarization becomes different for those two drain positions.
\subsubsection{Three components of the spin polarization coefficient: $P_x,P_y$, and $P_z$}
So far, we have discussed the spin-dependent transmission coefficients, currents and spin-polarized coefficient, where the spin-quantization axis was along the $z$-direction. Since the spin-sublattice symmetry is broken by the noncolinear arrangement of the magnetic moments in the SPG network, the spin-splitting effect should also occur for the other spin-quantization axes, namely, along the $x$ and $y$ directions. 
\begin{figure}[ht!]
\centering
\includegraphics[width=0.45\textwidth,height=0.3\textwidth]{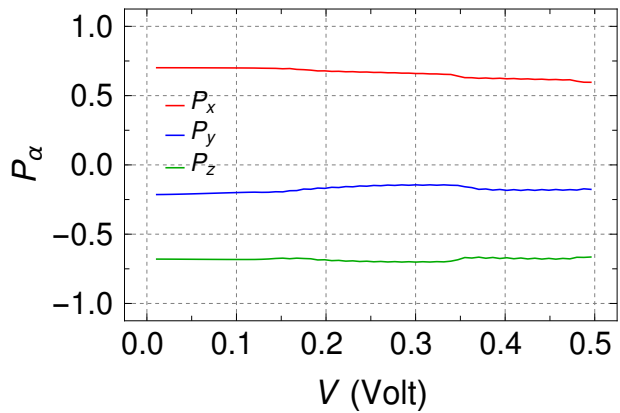} 
\caption{(Color online).  $P_\alpha$ $(\alpha=x,y,z)$ as a function of bias voltage for three different drain positions for a 5th generation SPG network. The blue, red, and green colors correspond to the results for $P_x$, $P_y$, and $P_z$, respectively.  Here, the chosen light parameters are $A_x =2.5$, $A_y = 0$ and the Fermi energy is set at $E_F =0.5$. Since for $A_y=0$, there is no specific role of $\phi$ in renormalizing the NNH integrals, we set $\phi=0$.}
\label{pxyz}
\end{figure}
Therefore, it is also important to study the behavior of the spin polarization coefficient for the other spin-quantization axes in presence of light. In Fig.~\ref{pxyz}, we show the behavior of the $x$, $y$, and $z$-components of the spin polarization coefficient as a function of the bias voltage. We consider a 5th generation SPG to compute $P_x$, $P_y$, and $P_z$. The light parameters are $A_x =2.5$ and $A_y = 0$. We fixed the Fermi energy at $E_F=0.5$. The $x$ and $z$-components of the spin polarization coefficients show favorable response, where the degree of spin polarization is about $70\%$, while the $y$-component shows a poor response for the particular choice of the light parameters. However, with appropriate choices of the light parameters and the Fermi energy, the degree of spin polarization of the $y$-component can also be enhanced. Overall, all the three components of the spin polarization show finite response, which certainly makes the present work more intriguing for spin-based applications. Experimentally, all the three components of spin polarization can be measured by using a Wien filter and Mott detector~\cite{pol-detection}.

\section{Concluding remarks}

In conclusion, we propose a new prescription to get significant spin polarization by irradiating an AFM SPG network. The noncolinear texture of the magnetic moments breaks spin-sublattice symmetry, yielding a non-zero spin polarization. We must mention that instead of modifying the physical system parameters, a high degree of spin polarization can be achieved just by irradiating the system, which is quite a significant achievement. To the best of our knowledge, such a prescription for spin-resolved transmission has not been reported so far in fractal structures. The system under investigation has been described within a tight-binding framework, and the irradiation effect has been incorporated using the Floquet-Bloch ansatz following the minimal coupling scheme. The spin-resolved transmission coefficients have been evaluated using the standard Green's function formalism based on Landauer-B\"{u}ttiker approach. Keeping in mind the localization phenomenon, at first, a higher generation (8th) SPG has been studied to examine the spin polarization in the presence and the absence of light. The effects of various physical parameters on spin polarization have been investigated thoroughly to make the present communication coherent and complete. Our essential findings and the important aspects of the present communication are summarized as follows.

$\bullet$ The noncolinear texture of the magnetic moments destroys the fractal nature of the energy spectra and splits the up and down spin transmission coefficients differently.

$\bullet$ We have observe the signature of multiple spin-dependent mobility edges both in the absence and presence of light.

$\bullet$ We have achieved a significant spin polarization for the noncolinear AFM SPG in the presence of light.

$\bullet$ The maximum spin polarization is obtained for the spin-dependent scattering parameter $h \sim 0.5\,$eV and shows uniform response for different generations of SPG.

$\bullet$ The magnitude of spin currents can significantly be enhanced (more than $90\%$) by irradiating the system with suitable light parameters.

$\bullet$ We have found a large degree of spin polarization over a wide range of irradiation parameters.

$\bullet$ The degree of spin polarization strongly depends on the drain position. 

$\bullet$ $x$, $y$, and $z$, all the three components of the spin polarization are non-zero due to the broken spin-sublattice symmetry and with a proper choice of the irradiation parameters, the degree of polarization can be enhanced significantly.

At the end, as the results reported in the present work, provide several important features of spin-dependent transport phenomena in a driven noncolinear AFM SPG network, such a scheme will surely attract to study the spintronics properties in other fractal structures and we may experience some novel features. The spectral peculiarity, fractal-like gapped energy spectrum, and the coexistence of both conducting and localized states play the central role of getting non-trivial signatures in antiferromagnetic fractal lattices. These features are no longer observed in completely perfect or fully uncorrelated (random) disordered lattices. Our analysis might help in designing efficient spin-based devices at the nanoscale in near future. 


\setcounter{secnumdepth}{0}
\section{ACKNOWLEDGMENTS}

KM and SG thank S. Chattopadhyay for the numerical facility at IITG. SKM thankfully acknowledges the financial support of the Science and Engineering Research Board, 
Department of Science and Technology, Government of India (Project File Number: EMR/2017/000504).


\end{document}